\documentclass{svjour2}                    
\smartqed  
\usepackage{graphicx}
\usepackage{amsmath}
\usepackage{amssymb}
\journalname{Celestial Mechanics and Dynamical Astronomy}
\begin{document}

\title{Figure-Figure Interaction Between Bodies Having Arbitrary Shapes and Mass Distributions: A Power Series Expansion Approach}



\author{Pasquale Tricarico}

\authorrunning{P.~Tricarico}

\institute{P.~Tricarico \at
           Planetary Science Institute \\
           1700 E.~Ft.~Lowell Rd., Suite 106, Tucson AZ 85719 \\
           Tel.: +001-520-622-6300\\
           Fax:  +001-520-622-8060\\
           \email{tricaric@psi.edu}
}

\date{Received: date / Accepted: date}

\maketitle

\begin{abstract}
We derive an expression for the mutual gravitational force and torque
of two bodies having arbitrary shapes and mass distributions,
as an expansion in power series of their products of inertia
and of the relative coordinates of their centres of mass.
The absolute convergence of all the power series developed is rigorously demonstrated.
The absence of transcendental functions makes this formalism suitable for fast numerical applications.
The products of inertia used here are directly related to the spherical harmonics coefficients, 
and we provide a detailed analysis of this relationship.
\end{abstract}

\section{Introduction}

The mutual gravitational potential of two physical bodies
contains the information necessary and sufficient 
to derive their mutual force and torque,
and thus completely describe their dynamical evolution.
However, in general the direct derivation of the mutual force and torque
from the mutual potential
can pose substantial mathematical difficulties.
In the literature, 
these difficulties have often been alleviated by making assumptions 
on the physical properties of the bodies
or on their dynamical configuration,
but of course this comes at the cost of the loss of generality.

In \cite{1978CeMec..18..295B}
the mutual gravitational potential of $N$ solid bodies is
expressed using the spherical harmonics coefficients of each body,
but in a formalism based on complex variables 
for the sake of a compact notation.
The expressions make use of transcendental functions,
that are computationally expensive and thus not prone to fast numerical applications.
No explicit expression for the mutual force and torque is provided.

A different approach is followed in \cite{2005CeMDA..91..337W} and \cite{2006CeMDA..96..317F},
where the shape of each body is described by a polyhedron,
and the spatial resolution of the mass density distribution is limited by the number of simplices.
Transcendental functions are avoided, at the benefit of fast numerical applications,
and the limitations imposed by the choice to model the shape of each body
with a polyhedron can be made arbitrarily small by increasing the number of vertexes,
at the expense of computational efficiency.

A basic but powerful approach is used in \cite{1988CeMec..44...49P}, 
where the mutual gravitational potential
is expressed in a formalism based on cartesian
coordinates and products of inertia,
with no assumptions on the shape and mass density of the bodies,
and avoiding transcendental functions.
This is the most generic formalism we could find in the literature.

In this work,
we choose an approach that allows to treat the figure-figure interaction of rigid bodies in its most generic form 
(arbitrary shape and mass distribution).
Given our previous analysis, the natural choice is to follow \cite{1988CeMec..44...49P}
and determine explicit expressions for the mutual gravitational force and torque.
We also improve on \cite{1988CeMec..44...49P} by demonstrating the absolute convergence
of all the power series developed, and by finding an explicit expression for $t_{l m n L M N}$, the main coefficient of the formalism, 
that was only defined recursively in \cite{1988CeMec..44...49P}.

The plan of the manuscript is the following:
in Sec.~(\ref{sec_grav_pot}) we introduce the main formalism that will
be used throughout this work;
in Sec.~(\ref{sec_ft}) we derive the complete expression for the mutual 
gravitational force and torque;
in Sec.~(\ref{sec_convergence}) we demonstrate the absolute convergence of all the series developed.
Two appendix sections further enrich this manuscript:
in Sec.~(\ref{sec_conversion}) we derive the conversion formulas 
between the products of inertia coefficients and the spherical harmonics coefficients;
in Sec.~(\ref{sec_precomputed}) we provide formulas for the 
products of inertia relative to bodies with a regular shape and an uniform density.

\section{Mutual Gravitational Potential\label{sec_grav_pot}}

In this section we introduce the formalism developed in \cite{1988CeMec..44...49P},
with some modifications and extensions.
Consider two bodies $b=1,2$ with centers of mass $\vec{O_b}$ and volumes $V_b$.
With respect to the inertial reference system, 
$\vec{O_1}=[0,0,0]^{\mathsf T}$ is chosen to be at the origin,
while $\vec{O_2}$ is located at $[\xi,\eta,\zeta]^{\mathsf T}$,
and the position of a generic point $P_b$ within $V_b$
is $\vec{P_1}=[X_1,Y_1,Z_1]^{\mathsf T}$
and $\vec{P_2}=[\xi+X_2,\eta+Y_2,\zeta+Z_2]^{\mathsf T}$.
The symbol ``$^{\mathsf T}$'' denotes the transpose operator.
In the body-fixed reference systems, 
we have $\vec{O_1}=[0,0,0]^{\mathsf T}$, $\vec{P_1}=[x_1,y_1,z_1]^{\mathsf T}$,
and $\vec{O_2}=[0,0,0]^{\mathsf T}$, $\vec{P_2}=[x_2,y_2,z_2]^{\mathsf T}$.
A vector in the inertial frame can be transformed
into corresponding vector in the body-fixed frame
through a rotation ${\cal Q}_b$:
\begin{eqnarray}
{\cal Q}_b &=&
\left[
\begin{array}{rrr}
l_{1b} & l_{2b} & l_{3b} \\
m_{1b} & m_{2b} & m_{3b} \\
n_{1b} & n_{2b} & n_{3b} 
\end{array}
\right]
\nonumber
\end{eqnarray}
such that $[x_b,y_b,z_b]^{\mathsf T} = {\cal Q}_b [X_b,Y_b,Z_b]^{\mathsf T}$.
Similarly, we define the vector 
$\vec{R} = \vec{O_2} - \vec{O_1} = [\xi,\eta,\zeta]^{\mathsf T}$ in the inertial frame,
that transforms as 
$[\xi_b,\eta_b,\zeta_b]^{\mathsf T} = {\cal Q}_b [\xi,\eta,\zeta]^{\mathsf T}$,
so that $R^2 = \xi^2 + \eta^2 + \zeta^2 = \xi_b^2 + \eta_b^2 + \zeta_b^2$.
The inverse of the distance $\rho$ between $\vec{P_1}$ and $\vec{P_2}$
can now be expressed using Eq.~(6$''$) in \cite{1988CeMec..44...49P}:
\begin{eqnarray}
\frac{1}{\rho} &=& \frac{1}{R}
\sum_{i_1=0}^{\infty} \sum_{j_1=0}^{\infty} \sum_{k_1=0}^{\infty}
\sum_{i_2=0}^{\infty} \sum_{j_2=0}^{\infty} \sum_{k_2=0}^{\infty}
(-1)^{i_1+j_1+k_1}
\frac{x_1^{i_1}y_1^{j_1}z_1^{k_1}x_2^{i_2}y_2^{j_2}z_2^{k_2}}{i_1!j_1!k_1!i_2!j_2!k_2!}
\times \nonumber \\
&\times& 
\sum_{i_3=0}^{i_2} \sum_{j_3=0}^{j_2} \sum_{k_3=0}^{k_2} 
\sum_{i_4=0}^{i_3} \sum_{j_4=0}^{j_3} \sum_{k_4=0}^{k_3}
{i_2 \choose i_3} {j_2 \choose j_3} {k_2 \choose k_3} 
{i_3 \choose i_4} {j_3 \choose j_4} {k_3 \choose k_4} 
\times \nonumber \\
&\times& 
l_x^{i_4} m_x^{i_3-i_4} n_x^{i_2-i_3}
l_y^{j_4} m_y^{j_3-j_4} n_y^{j_2-j_3}
l_z^{k_4} m_z^{k_3-k_4} n_z^{k_2-k_3}
\times \nonumber \\
&\times& 
\sum_{L=0}^{i_5} \sum_{M=0}^{j_5} \sum_{N=0}^{k_5}
t_{i_5 j_5 k_5 L M N}
\frac{\xi_1^L \eta_1^M \zeta_1^N}{R^{i_1 + j_1 + k_1 + i_2 + j_2 + k_2 + L + M + N}}
\label{eq_one_over_rho}
\end{eqnarray}
where
\begin{eqnarray}
i_5 &=& i_1 + i_4 + j_4 + k_4 \nonumber \\
j_5 &=& j_1 + i_3 - i_4 + j_3 - j_4 + k_3 - k_4 \nonumber \\
k_5 &=& k_1 + i_2 - i_3 + j_2 - j_3 + k_2 - k_3 \nonumber
\end{eqnarray}
and $i_5+j_5+k_5=i_1+i_2+j_1+j_2+k_1+k_2$.
The coefficient $t_{l m n L M N}$ is defined by the recursive relation
\begin{eqnarray}
t_{l m n L M N} 
&=& \left\{ (2l-C_{lmn}) t_{(l-1)mn(L-1)MN} \right. + \nonumber \\
&& + \left. (l-1) (l-C_{lmn}) t_{(l-2)mnLMN}\right\} (\delta_{l0}-1) + \nonumber \\
&+& \left\{ (2m-C_{lmn}) t_{l(m-1)nL(M-1)N} \right. + \nonumber \\
&& + \left. (m-1) (m-C_{lmn}) t_{l(m-2)nLMN}\right\} (\delta_{m0}-1) + \nonumber \\
&+& \left\{ (2n-C_{lmn}) t_{lm(n-1)LM(N-1)} \right. + \nonumber \\
&& + \left. (n-1) (n-C_{lmn}) t_{lm(n-2)LMN}\right\} (\delta_{n0}-1)
\label{eq_recursive_t}
\end{eqnarray}
with $C_{lmn} = 1 / \left( 3 - \delta_{l0} - \delta_{m0} - \delta_{n0} \right)$,
and $\delta_{ij}$ the Kronecker delta function.
The initial value is $t_{0 0 0 0 0 0} \equiv 1$, and the coefficient is zero
when any one of $(l+L)$, $(m+M)$, or $(n+N)$ is an odd integer,
or when any of the six indices assumes a negative value.
A table containing the coefficients up to $l+m+n \leq 7$
is available in \cite{1988CeMec..44...49P},
and in Eq.~(\ref{eq_explicit_t}) we derive an explicit formula for $t_{l m n L M N}$.
Finally, the coefficients $l_x, m_x, \dots, n_z$ are defined by:
\begin{eqnarray}
\left[
\begin{array}{rrr}
l_{x} & l_{y} & l_{z} \\
m_{x} & m_{y} & m_{z} \\
n_{x} & n_{y} & n_{z} 
\end{array}
\right] &=& 
{\cal Q}_1 {\cal Q}_2^{\mathsf T}
\label{eq_QQ}
\end{eqnarray}
where the matrix ${\cal Q}_1 {\cal Q}_2^{\mathsf T}$ clearly represents the relative rotation between the two bodies.

The mutual gravitational potential $U_{12}$ can now be written as
\begin{equation}
\displaystyle
U_{12} = G \int_{V_1} \int_{V_2} \frac{\delta_1(x_1,y_1,z_1) \ \delta_2(x_2,y_2,z_2)}{\rho} \ \mbox{d}V_1 \mbox{d}V_2
\label{eq_potential}
\end{equation}
where $G$ is the gravitational constant and
$\delta_b(x_b,y_b,z_b)$ is the mass density of the body $b$.
By substituting Eq.~(\ref{eq_one_over_rho}) in Eq.~(\ref{eq_potential}),
we obtain integrals defined as 
\emph{generalized products of inertia} in \cite{1988CeMec..44...49P},
of the form
\begin{equation}
{\cal M}_{b,ijk} = \int_{V_b} x_b^i y_b^j z_b^k \delta_b(x_b,y_b,z_b) \ \mbox{d}V_b
\label{eq_def_M}
\end{equation}
and since these integrals involve only coordinates relative to the body $b$ 
in the body-fixed reference frame,
they play the same role of the $\{C_{lm},S_{lm}\}$ spherical harmonics coefficients (see Sec.~(\ref{sec_conversion})),
and need to be computed only once for a body with constant 
shape and mass distribution.
In this manuscript we prefer to deal with adimensional coefficients,
so we introduce the \emph{normalized products of inertia}
${\cal N}_{b,ijk}$ defined by
\begin{equation}
\displaystyle
{\cal N}_{b,ijk} \equiv
\frac{\displaystyle \int_{V_b} \frac{x_b^i y_b^j z_b^k}{r_0^{i+j+k}} \delta_b(x_b,y_b,z_b) \ {\mathrm d}V_b}{\displaystyle \int_{V_b} \delta_b(x_b,y_b,z_b) \ {\mathrm d}V_b} =
\frac{1}{r_0^{i+j+k}} \frac{{\cal M}_{b,ijk}}{{\cal M}_{b,000}}
\label{eq_def_N}
\end{equation}
where $r_0$ is an arbitrary normalization radius,
and ${\cal M}_{b,000} \equiv {\cal M}_b$ is simply the mass of the body $b$.
The final expression for the mutual gravitational potential is then:
\begin{eqnarray}
U_{12} &=& 
G 
\frac{{\cal M}_1 {\cal M}_2}{R}
\sum_{i_1=0}^{\infty} \sum_{j_1=0}^{\infty} \sum_{k_1=0}^{\infty}
\sum_{i_2=0}^{\infty} \sum_{j_2=0}^{\infty} \sum_{k_2=0}^{\infty}
(-1)^{i_1+j_1+k_1}
\times \nonumber \\
&\times& 
\frac{{\cal N}_{1,i_1 j_1 k_1}}{i_1!j_1!k_1!}
\frac{{\cal N}_{2,i_2 j_2 k_2}}{i_2!j_2!k_2!}
\left(\frac{r_0}{R}\right)^{i_1 + j_1 + k_1 + i_2 + j_2 + k_2}
\times \nonumber \\
&\times& 
\sum_{i_3=0}^{i_2} \sum_{j_3=0}^{j_2} \sum_{k_3=0}^{k_2} 
\sum_{i_4=0}^{i_3} \sum_{j_4=0}^{j_3} \sum_{k_4=0}^{k_3}
{i_2 \choose i_3} {j_2 \choose j_3} {k_2 \choose k_3} 
{i_3 \choose i_4} {j_3 \choose j_4} {k_3 \choose k_4} 
\times \nonumber \\
&\times& 
l_x^{i_4} m_x^{i_3-i_4} n_x^{i_2-i_3}
l_y^{j_4} m_y^{j_3-j_4} n_y^{j_2-j_3}
l_z^{k_4} m_z^{k_3-k_4} n_z^{k_2-k_3}
\times \nonumber \\
&\times& 
\sum_{L=0}^{i_5} \sum_{M=0}^{j_5} \sum_{N=0}^{k_5}
t_{i_5 j_5 k_5 L M N}
\frac{\xi_1^L \eta_1^M \zeta_1^N}{R^{L+M+N}}
\label{eq_potential_extended_final}
\end{eqnarray}
The infinite series appearing in these expressions 
are typically truncated in real-world applications.
As we show in Sec.~(\ref{sec_conversion}),
the order of the ${\cal N}_{ijk}$ coefficients is given by $i+j+k$.
The level of accuracy required by the particular problem being investigated
drives the order of the coefficients to be used.

\section{Mutual Gravitational Force and Torque\label{sec_ft}}

The expression
\begin{equation*}
-G\delta_1(P_1)\delta_2(P_2)\nabla(1/\rho) \ \mbox{d}V_1 \mbox{d}V_2
\end{equation*}
represents the force acting on $P_1$ 
due to the gravitational interaction between two infinitesimal masses with volume $\mbox{d}V_b$
and densities $\delta_b(P_b)$
located at $P_b$, with positive direction $P_2-P_1$.
In order to obtain the total force acting on the barycenter of body $1$, we need an expression for the gradient
$\nabla(1/\rho)$ with components:
\begin{eqnarray}
\nabla \left( {1}/{\rho} \right) &=&
\left[
\frac{\partial}{\partial\xi_1},
\frac{\partial}{\partial\eta_1},
\frac{\partial}{\partial\zeta_1}
\right]^{\mathsf T}
\frac{1}{\rho}
\nonumber
\end{eqnarray}
The only part of $1/\rho$
affected by the gradient is ${\xi_1^L \eta_1^M \zeta_1^N}/{R^{\wp+1}}$ (see Eq.~(\ref{eq_one_over_rho})),
where we have defined $\wp=i_1+j_1+k_1+i_2+j_2+k_2+L+M+N$ for brevity, and where
$R^2 = \xi^2 + \eta^2 + \zeta^2 = \xi_1^2 + \eta_1^2 + \zeta_1^2$.
Explicitly, the effect of the gradient on the affected part of $1/\rho$ is:
\begin{eqnarray}
\left[
\begin{array}{r}
{\partial}/{\partial\xi_1} \\
{\partial}/{\partial\eta_1} \\
{\partial}/{\partial\zeta_1}
\end{array}
\right]
\frac{\xi_1^L \eta_1^M \zeta_1^N}{R^{\wp+1}}
&=&
\frac{\xi_1^L \eta_1^M \zeta_1^N}{R^{\wp+1}}
\left\{
\left[ 
\begin{array}{r}
{L}/{\xi_1} \\
{M}/{\eta_1} \\
{N}/{\zeta_1}
\end{array}
\right]
- \frac{\wp+1}{R^2}
\left[
\begin{array}{r}
\xi_1 \\
\eta_1 \\
\zeta_1
\end{array}
\right]
\right\}
\nonumber
\end{eqnarray}
The total force $F_{12}$ acting on body $1$
is determined by integrating over the volume of the two bodies, obtaining:
\begin{eqnarray}
F_{12} &=& 
G 
\frac{{\cal M}_1 {\cal M}_2}{R}
\sum_{i_1=0}^{\infty} \sum_{j_1=0}^{\infty} \sum_{k_1=0}^{\infty}
\sum_{i_2=0}^{\infty} \sum_{j_2=0}^{\infty} \sum_{k_2=0}^{\infty}
(-1)^{i_1+j_1+k_1}
\times \nonumber \\
&\times& 
\frac{{\cal N}_{1,i_1 j_1 k_1}}{i_1!j_1!k_1!}
\frac{{\cal N}_{2,i_2 j_2 k_2}}{i_2!j_2!k_2!}
\left(\frac{r_0}{R}\right)^{i_1 + j_1 + k_1 + i_2 + j_2 + k_2}
\times \nonumber \\
&\times& 
\sum_{i_3=0}^{i_2} \sum_{j_3=0}^{j_2} \sum_{k_3=0}^{k_2} 
\sum_{i_4=0}^{i_3} \sum_{j_4=0}^{j_3} \sum_{k_4=0}^{k_3}
{i_2 \choose i_3} {j_2 \choose j_3} {k_2 \choose k_3} 
{i_3 \choose i_4} {j_3 \choose j_4} {k_3 \choose k_4} 
\times \nonumber \\
&\times& 
l_x^{i_4} m_x^{i_3-i_4} n_x^{i_2-i_3}
l_y^{j_4} m_y^{j_3-j_4} n_y^{j_2-j_3}
l_z^{k_4} m_z^{k_3-k_4} n_z^{k_2-k_3}
\times \nonumber \\
&\times& 
\sum_{L=0}^{i_5} \sum_{M=0}^{j_5} \sum_{N=0}^{k_5}
t_{i_5 j_5 k_5 L M N}
\frac{\xi_1^L \eta_1^M \zeta_1^N}{R^{L+M+N}}
\left\{
 \frac{\wp+1}{R^2}
\left[
\begin{array}{r}
\xi_1 \\
\eta_1 \\
\zeta_1
\end{array}
\right]
-\left[ 
\begin{array}{r}
{L}/{\xi_1} \\
{M}/{\eta_1} \\
{N}/{\zeta_1}
\end{array}
\right]
\right\}
\label{eq_force_extended_final}
\end{eqnarray}

In a similar fashion, the torque 
acting on the barycenter of body $1$ due to the mutual force between the same two infinitesimal masses
is given by the vector product 
\begin{equation*}
-G\delta_1(P_1)\delta_2(P_2)[x_1,y_1,z_1]^{\mathsf T}\wedge\nabla(1/\rho)\ \mbox{d}V_1 \mbox{d}V_2
\end{equation*}
and the total torque $\tau_{12}$ acting on body $1$ is therefore:
\begin{eqnarray}
\tau_{12} &=& 
G 
\frac{{\cal M}_1 {\cal M}_2}{R}
\sum_{i_1=0}^{\infty} \sum_{j_1=0}^{\infty} \sum_{k_1=0}^{\infty}
\sum_{i_2=0}^{\infty} \sum_{j_2=0}^{\infty} \sum_{k_2=0}^{\infty}
(-1)^{i_1+j_1+k_1}
\times \nonumber \\
&\times& 
\frac{1}{i_1!j_1!k_1!}
\frac{{\cal N}_{2,i_2 j_2 k_2}}{i_2!j_2!k_2!}
\left(\frac{r_0}{R}\right)^{i_1 + j_1 + k_1 + i_2 + j_2 + k_2}
\times \nonumber \\
&\times& 
\sum_{i_3=0}^{i_2} \sum_{j_3=0}^{j_2} \sum_{k_3=0}^{k_2} 
\sum_{i_4=0}^{i_3} \sum_{j_4=0}^{j_3} \sum_{k_4=0}^{k_3}
{i_2 \choose i_3} {j_2 \choose j_3} {k_2 \choose k_3} 
{i_3 \choose i_4} {j_3 \choose j_4} {k_3 \choose k_4} 
\times \nonumber \\
&\times& 
l_x^{i_4} m_x^{i_3-i_4} n_x^{i_2-i_3}
l_y^{j_4} m_y^{j_3-j_4} n_y^{j_2-j_3}
l_z^{k_4} m_z^{k_3-k_4} n_z^{k_2-k_3}
\times \nonumber \\
&\times& 
\sum_{L=0}^{i_5} \sum_{M=0}^{j_5} \sum_{N=0}^{k_5}
t_{i_5 j_5 k_5 L M N}
\frac{\xi_1^L \eta_1^M \zeta_1^N}{R^{L+M+N}}
\times \nonumber \\
&\times& 
\left\{
\frac{\wp+1}{R^2}
\left[
\begin{array}{c}
{\cal N}_{1,i_1,j_1+1,k_1} \zeta_1 - {\cal N}_{1,i_1,j_1,k_1+1} \eta_1 \\
{\cal N}_{1,i_1,j_1,k_1+1} \xi_1   - {\cal N}_{1,i_1+1,j_1,k_1} \zeta_1 \\
{\cal N}_{1,i_1+1,j_1,k_1} \eta_1  - {\cal N}_{1,i_1,j_1+1,k_1} \xi_1
\end{array}
\right]\right.
+ \nonumber \\
&& - \left.\left[
\begin{array}{c}
{\cal N}_{1,i_1,j_1+1,k_1} N/{\zeta_1} - {\cal N}_{1,i_1,j_1,k_1+1} M/{\eta_1} \\
{\cal N}_{1,i_1,j_1,k_1+1} L/{\xi_1}   - {\cal N}_{1,i_1+1,j_1,k_1} N/{\zeta_1} \\
{\cal N}_{1,i_1+1,j_1,k_1} M/{\eta_1}  - {\cal N}_{1,i_1,j_1+1,k_1} L/{\xi_1}
\end{array}
\right]
\right\}
\label{eq_torque_extended_final}
\end{eqnarray}
Both Eq.~(\ref{eq_force_extended_final},\ref{eq_torque_extended_final})
are expressed in the reference system relative to body $1$.
It is important to stress that the only singularity
in Eq.~(\ref{eq_force_extended_final},\ref{eq_torque_extended_final})
is still in $R \rightarrow 0$, as a possibly suspicious term of the type $L/{\xi_1}$
will turn out to be regular when each equation is expanded and the term
becomes of the type $L \xi_1^{L-1}$.

\section{Formal Convergence\label{sec_convergence}}

In this section, we analyze the convergence properties 
of the power series expansions contained in this manuscript.
We start with a basic relation 
developed in section 202 of \cite{potential.book},
for the vectors $\vec \rho = \vec R - \vec r$:
\begin{eqnarray}
\frac{1}{\rho} &=&
\frac{1}{R}
\sum_{n=0}^{\infty} 
\left(\frac{r}{R}\right)^n 
P_n(\cos\lambda)
\label{eq_base_libro}
\end{eqnarray}
with $r=\sqrt{x^2+y^2+z^2}$, 
$P_n(\cos\lambda)$ the Legendre polynomial of $\cos\lambda$,
and $\lambda$ the angle between the two vectors $\vec R$ and $\vec r$.
The power series in Eq.~(\ref{eq_base_libro}) converges absolutely,
because $|P_n(\cos\lambda)| \leq 1$ for any integer $n \geq 0$ and for any angle $\lambda$, and:
\begin{eqnarray}
\sum_{n=0}^{\infty} 
\left|
\left(\frac{r}{R}\right)^n 
P_n(\cos\lambda)
\right|
\leq 
\sum_{n=0}^{\infty} 
\left(\frac{r}{R}\right)^n 
=
\left(1-\frac{r}{R}\right)^{-1}
\nonumber
\end{eqnarray}
for $r<R$, where both $r$ and $R$ are positive.
This expression for $1/\rho$ 
can be further expanded while conserving the same convergence characteristics, 
and section 204 of \cite{potential.book} gives:
\begin{eqnarray}
\frac{1}{\rho} 
&=&
\frac{1}{R}
\sum_{n=0}^{\infty} 
\sum_{s=0}^{\lfloor n/2 \rfloor}
\left(-\frac{1}{2}\right)^s
(2n-2s-1)!!
\sum_{\substack{i, j, k \\ i+j+k=n}} 
\frac{x^i y^j z^k}{R^{n}}
\times \nonumber \\
&\times&
\sum_{\substack{\alpha, \beta, \gamma \\ \alpha+\beta+\gamma=s}} 
\frac{1}{\alpha! \beta! \gamma!}
\frac{1}{(i-2\alpha)!(j-2\beta)!(k-2\gamma)!}
\frac{\xi^{i-2\alpha} \eta^{j-2\beta} \zeta^{k-2\gamma}}{R^{n-2s}}
\nonumber
\end{eqnarray}
If we substitute 
$\vec r = [x,y,z]^{\mathsf T}$ 
with $\vec{r_2} - \vec{r_1} = [x_2-x_1,y_2-y_1,z_2-z_1]^{\mathsf T}$
and then apply the binomial theorem, we obtain:
\begin{eqnarray}
\frac{1}{\rho} 
&=&
\frac{1}{R}
\sum_{n=0}^{\infty} 
\sum_{s=0}^{\lfloor n/2 \rfloor}
\left(-\frac{1}{2}\right)^s
(2n-2s-1)!!
\times \nonumber \\
&\times&
\sum_{\substack{i, j, k \\ i+j+k=n}} 
\sum_{a=0}^i \sum_{b=0}^j \sum_{c=0}^k 
{i \choose a} {j \choose b} {k \choose c} 
(-1)^{a+b+c} 
\frac{
x_1^{i-a}
y_1^{j-b}
z_1^{k-c}
x_2^a 
y_2^b 
z_2^c}{R^n}
\times \nonumber \\
&\times&
\sum_{\substack{\alpha, \beta, \gamma \\ \alpha+\beta+\gamma=s}} 
\frac{1}{\alpha! \beta! \gamma!}
\frac{1}{(i-2\alpha)!(j-2\beta)!(k-2\gamma)!}
\frac{\xi^{i-2\alpha} \eta^{j-2\beta} \zeta^{k-2\gamma}}{R^{n-2s}}
\label{eq_book_expanded.three}
\end{eqnarray}
In order to compare Eq.~(\ref{eq_book_expanded.three}) with Eq.~(\ref{eq_one_over_rho}),
we assume for a moment no relative rotation between 
the two body-fixed reference frames in Eq.~(\ref{eq_one_over_rho}), so that the product 
${\cal Q}_1 {\cal Q}_2^{\mathsf T}$ is equal to the identity matrix.
This leads to the following simplified expression for Eq.~(\ref{eq_one_over_rho}):
\begin{eqnarray}
\frac{1}{\rho} &=& \frac{1}{R}
\sum_{i_1=0}^{\infty} \sum_{j_1=0}^{\infty} \sum_{k_1=0}^{\infty}
\sum_{i_2=0}^{\infty} \sum_{j_2=0}^{\infty} \sum_{k_2=0}^{\infty}
(-1)^{i_1+j_1+k_1}
\frac{x_1^{i_1}y_1^{j_1}z_1^{k_1}x_2^{i_2}y_2^{j_2}z_2^{k_2}}{i_1!j_1!k_1!i_2!j_2!k_2!}
\times \nonumber \\
&\times& 
\sum_{L=0}^{i_5} \sum_{M=0}^{j_5} \sum_{N=0}^{k_5}
t_{i_5 j_5 k_5 L M N}
\frac{\xi_1^L \eta_1^M \zeta_1^N}{R^{i_1 + j_1 + k_1 + i_2 + j_2 + k_2 + L + M + N}}
\label{eq_rho.identity}
\end{eqnarray}
where we have now
$i_5=i_1+i_2$,
$j_5=j_1+j_2$,
$k_5=k_1+k_2$.
The expressions in Eq.~(\ref{eq_book_expanded.three}) and Eq.~(\ref{eq_rho.identity})
can now be compared:
the fact that both equations represent the same quantity $1/\rho$
is not guarantee that they have the same convergence characteristics.
But if we can demonstrate that one equation is nothing more than an algebraic
manipulation of the other, then the same convergence of both equations is guaranteed.
This is the case for Eq.~(\ref{eq_book_expanded.three}) 
and Eq.~(\ref{eq_rho.identity}),
because we can obtain an explicit expression 
for the coefficient that was originally defined recursively in Eq.~(\ref{eq_recursive_t})
by comparing the two equations:
\begin{eqnarray}
t_{i j k L M N} &=& 
(-1)^{n-2s}
\left(-\frac{1}{2}\right)^s
\frac{i!j!k!}{\alpha! \beta! \gamma!}
\frac{(2n-2s-1)!!}{(i-2\alpha)!(j-2\beta)!(k-2\gamma)!}
\nonumber \\ 
&=&
\frac{\displaystyle(-1)^{(s_{ijk}+S_{LMN})/2}}{\displaystyle2^{(s_{ijk}-S_{LMN})/2}}
\frac{i!j!k!}{\alpha!\beta!\gamma!}
\frac{(s_{ijk}+S_{LMN}-1)!!}{L!M!N!}
\label{eq_explicit_t}
\end{eqnarray}
where $n=s_{ijk}=i+j+k$, $S_{LMN}=L+M+N$, $2s=n-S_{LMN}$, 
$L=i-2\alpha$, $M=j-2\beta$, $N=k-2\gamma$, 
$s=\alpha+\beta+\gamma$.
This remarkable result, expressing the key coefficient of Eq.~(\ref{eq_rho.identity})
using coefficients in Eq.~(\ref{eq_book_expanded.three}), 
clearly demonstrates that the two series are identical, 
and their different appearance is only due to elementary algebraic manipulation.
We conclude that Eq.~(\ref{eq_rho.identity})
converges absolutely for $|\vec{r_2} - \vec{r_1}| < R$.
In the general case, when a relative rotation between the two body-fixed reference frames
is present, the expression for $1/\rho$ in Eq.~(\ref{eq_one_over_rho}) still converges, 
as we didn't assume any particular value for 
the variables $x_b$, $y_b$, $z_b$ and $\xi$, $\eta$, $\zeta$.
But in this case the convergence condition becomes
$|\vec{r_2} + \vec{r_1}| < R$,
the \emph{worst case} for the non-rotated convergence condition.

The expression for the gravitational potential $U_{12}$ in Eq.~(\ref{eq_potential_extended_final}),
as defined in Eq.~(\ref{eq_potential}), converges where the expression for $1/\rho$ converges,
that is, in every point outside the two bounding spheres relative to the two bodies,
as long as the two spheres have no points in common.

Finally, the expressions for 
the force $F_{12}$ in Eq.~(\ref{eq_force_extended_final})
and the torque $\tau_{12}$ in Eq.~(\ref{eq_torque_extended_final})
both involve the gradient of $1/\rho$,
that can now be derived from Eq.~(\ref{eq_base_libro})
using the relation $(1-x^2) P_n'(x) = n P_{n-1}(x) - n x P_n(x)$
and the expansion $\cos\lambda = (x\xi+y\eta+z\zeta)/(r R)$.
The resulting expression for the gradient of $1/\rho$
is composed by three infinite series that can be generalized by:
\begin{eqnarray}
\sum_{n=0}^{\infty} \left(\frac{r}{R}\right)^n (n+p) P_{n-q}(\cos\lambda) \nonumber
\end{eqnarray}
with $p$ and $q$ integers equal to $0$ or $1$.
This series converges absolutely:
\begin{eqnarray}
\sum_{n=0}^{\infty} 
\left|
\left(\frac{r}{R}\right)^n 
(n+p) 
P_{n-q}(\cos\lambda)
\right|
\leq 
\sum_{n=0}^{\infty} 
\left(\frac{r}{R}\right)^n 
(n+p) 
=
\frac{(1-p) r R + p R^2}{(R-r)^2}
\nonumber
\end{eqnarray}
for $r<R$, where both $r$ and $R$ are positive.
This completes the demonstration that both Eq.~(\ref{eq_force_extended_final})
and Eq.~(\ref{eq_torque_extended_final}) converge absolutely.

\section{Conclusions\label{sec_conclusions}}

This manuscript provides the equations necessary to apply the
formalism initially developed by \cite{1988CeMec..44...49P}
to dynamical problems involving the gravitational interaction of 
an arbitrary number of bodies with arbitrary shape and mass distribution.
The convergence requirements are that any two 
bounding spheres relative to two bodies never intersect,
and this is analogous to formalisms based on 
a spherical harmonics expansion of the gravitational potential.
The dynamical evolution of an arbitrary number of solid bodies 
with arbitrary shape and mass distribution
is completely determined by the expressions for the
mutual force and torque provided in Sec.~(\ref{sec_ft}).
The products of inertia relative to each body 
can be obtained via direct integration (Eq.~(\ref{eq_def_M},\ref{eq_def_N})),
and in special cases, for homogeneous bodies 
with the shape of a box, a cylinder, or a triaxial ellipsoid, 
the products of inertia have been computed up to any order
and the results are presented in Sec.~(\ref{sec_precomputed}).
The connection between spherical harmonics coefficients 
and products of inertia coefficients 
is analyzed in great detail in Sec.~(\ref{sec_conversion}),
that provides explicit relations between the two.
Finally, in Sec.~(\ref{sec_convergence}) we 
demonstrate the absolute convergence 
of all the power series developed in the manuscript,
and in Eq.~(\ref{eq_explicit_t}) derive an explicit expression 
for the $t_{l m n L M N}$
coefficient, that plays a fundamental role in this formalism.



\appendix

\section{Conversion Between ${\cal N}_{ijk}$ and $\{C_{lm},S_{lm}\}$ Coefficients \label{sec_conversion}}

In this section, we derive the conversion formulas between the 
spherical harmonics coefficients $\{C_{lm},S_{lm}\}$ and ${\cal N}_{ijk}$,
for a body having arbitrary shape and mass distribution.
The spherical harmonics coefficients are typically defined by:
\begin{eqnarray}
\displaystyle
C_{lm} &=& \frac{1}{{\cal M}} \displaystyle \int_V (r/r_0)^l P_{lm}(z/r) \cos(m\phi) \delta(r) \ \mbox{d}V \label{eq_def_Clm} \\
S_{lm} &=& \frac{1}{{\cal M}} \displaystyle \int_V (r/r_0)^l P_{lm}(z/r) \sin(m\phi) \delta(r) \ \mbox{d}V \label{eq_def_Slm}
\end{eqnarray}
where ${\cal M}$ is the mass of the body,
$r=\sqrt{x^2+y^2+z^2}$,
$r_0$ is an arbitrary normalization radius,
$P_{lm}(z/r)$ is the associated Legendre function of $z/r$,
$\tan \phi = y/x$,
$\delta(r)$ is the mass density at $r$,
and $V$ is the body's volume.
With the intent to use a light notation, 
the coefficients $C_{lm}$ and $S_{lm}$ defined above do not include any normalization factor.
Any normalization factor can be easily added at any point in this section.

In order to proceed, we need to transform the expressions in Eq.~(\ref{eq_def_Clm},\ref{eq_def_Slm})
to make explicit the dependence on the coordinates $x$, $y$, $z$.
Consider the expansion for the associated Legendre function:
\begin{eqnarray}
P_l^{m}(w) &=& 
(-1)^m (1 - w^2)^{m/2} 2^{-l} \times \nonumber \\
&\times& 
\sum_{k=0}^{\lfloor l/2 \rfloor} 
(-1)^k {l \choose k} {2 l - 2 k \choose l} 
(l - m - 2 k + 1)_{m} w^{l - m - 2 k} \nonumber
\end{eqnarray}
where the notation $\lfloor a \rfloor$ represents the floor of $a$,
and $(a)_{m}$ is the Pochhammer function of $a$.
Since in the literature the associated Legendre function is defined up to a 
factor $(-1)^{m}$, here we will use the notation $P_{lm} (w) = (-1)^{m} P_l^{m}(w)$.
We use expansion of $\cos(m \phi)$ and $\sin(m \phi)$: 
\begin{eqnarray}
\cos(m \phi) &=& \sum_{k=0}^{\lfloor m/2 \rfloor} 
(-1)^k {m \choose 2 k} 
\sin^{2k}(\phi) 
\cos^{m - 2 k}(\phi) 
\nonumber \\
\sin(m \phi) &=& \sum_{k=0}^{\lfloor (m-1)/2 \rfloor}   
(-1)^k {m \choose 2 k + 1} 
\sin^{2 k + 1}(\phi)
\cos^{m - 2 k - 1}(\phi)
\nonumber
\end{eqnarray}
As we have 
$\sqrt{1 - (z/r)^2} \cos(\phi) = x/r$ 
and
$\sqrt{1 - (z/r)^2} \sin(\phi) = y/r$ 
we can now write:
\begin{eqnarray}
&& (r/r_0)^l P_{lm}(z/r) \cos(m\phi) =
2^{-l}
\sum_{p=0}^{\lfloor l/2 \rfloor} 
\sum_{q=0}^{\lfloor m/2 \rfloor} 
(-1)^{p+q}
{l \choose p} 
{2 l - 2 p \choose l} \times \nonumber \\
&& \ \ \ \ \ \ \times
{m \choose 2 q} (l - m - 2 p + 1)_{m}
x^{m - 2 q} y^{2 q} z^{l - m - 2 p} r^{2p} / r_0^l 
\nonumber
\\
&& (r/r_0)^l P_{lm}(z/r) \sin(m\phi) =
2^{-l}
\sum_{p=0}^{\lfloor l/2 \rfloor} 
\sum_{q=0}^{\lfloor (m-1)/2 \rfloor}   
(-1)^{p+q}
{l \choose p} 
{2 l - 2 p \choose l}  \times \nonumber \\
&& \ \ \ \ \ \ \times
{m \choose 2 q + 1} 
(l - m - 2 p + 1)_{m} 
x^{m - 2 q - 1}
y^{2 q + 1} 
z^{l - m - 2 p} 
r^{2p} / r_0^l
\nonumber
\end{eqnarray}
But before performing the integrals in Eq.~(\ref{eq_def_Clm},\ref{eq_def_Slm}) 
we need to expand $r^{2p} = \left( x^2 + y^2 + z^2 \right)^p$ as
\begin{eqnarray}
r^{2p} =
\sum_{\nu_x=0}^p
\sum_{\nu_y=0}^{p-\nu_x}
\frac{p!}{\nu_x!\nu_y!(p-\nu_x-\nu_y)!}
x^{2\nu_x} y^{2\nu_y} z^{2p-2\nu_x-2\nu_y}
\nonumber
\end{eqnarray}
and we finally obtain:
\begin{eqnarray}
\label{C_series_final}
C_{lm} &=&
2^{-l}
\sum_{p=0}^{\lfloor l/2 \rfloor} 
\sum_{q=0}^{\lfloor m/2 \rfloor} 
(-1)^{p+q}
{l \choose p} 
{2 l - 2 p \choose l} 
{m \choose 2 q} \times \nonumber \\
&\times& 
(l - m - 2 p + 1)_{m}
\sum_{\nu_x=0}^p
\sum_{\nu_y=0}^{p-\nu_x}
\frac{p!}{\nu_x!\nu_y!(p-\nu_x-\nu_y)!}
\times \nonumber \\
&\times& 
{\cal N}_{m-2q+2\nu_x,2q+2\nu_y,l-m-2\nu_x-2\nu_y} \\
\label{S_series_final}
S_{lm} &=&
2^{-l}
\sum_{p=0}^{\lfloor l/2 \rfloor} 
\sum_{q=0}^{\lfloor (m-1)/2 \rfloor}   
(-1)^{p+q}
{l \choose p} 
{2 l - 2 p \choose l} 
{m \choose 2 q + 1}  \times \nonumber \\
&\times& 
(l - m - 2 p + 1)_{m} 
\sum_{\nu_x=0}^p
\sum_{\nu_y=0}^{p-\nu_x}
\frac{p!}{\nu_x!\nu_y!(p-\nu_x-\nu_y)!}
\times \nonumber \\
&\times& 
{\cal N}_{m-2q-1+2\nu_x,2q+1+2\nu_y,l-m-2\nu_x-2\nu_y}
\end{eqnarray}
These two equations solve the direct problem of 
expressing the spherical harmonics coefficients $\{C_{lm},S_{lm}\}$
as a function of the normalized products of inertia ${\cal N}_{ijk}$.
The terms up to the fourth order for $C_{lm}$
(omitting the trivial values $C_{00} = {\cal N}_{000} = 1$, $C_{10} = {\cal N}_{001} = 0$, $C_{11} = {\cal N}_{100} = 0$,
$S_{11} = {\cal N}_{010} = 0$ when the center of mass is at the origin) are:
\begin{eqnarray}
C_{20} &=& {\cal N}_{002} - 1/2 \left( {\cal N}_{020} + {\cal N}_{200} \right) \nonumber \\
C_{21} &=& 3 {\cal N}_{101} \nonumber \\
C_{22} &=& 3 \left( {\cal N}_{200} - {\cal N}_{020} \right) \nonumber \\
C_{30} &=& {\cal N}_{003} - 3/2 \left( {\cal N}_{021} + {\cal N}_{201} \right) \nonumber \\
C_{31} &=& 6 {\cal N}_{102} - 3/2 \left( {\cal N}_{120}  + {\cal N}_{300} \right) \nonumber \\
C_{32} &=& 15 \left( {\cal N}_{201} - {\cal N}_{021} \right) \nonumber \\
C_{33} &=& 15 {\cal N}_{300} - 45 {\cal N}_{120} \nonumber \\
C_{40} &=& {\cal N}_{004} - 3 \left( {\cal N}_{202} + {\cal N}_{022} \right) + 3/4 {\cal N}_{220} + 3/8 \left( {\cal N}_{040} + {\cal N}_{400} \right) \nonumber \\
C_{41} &=& 10 {\cal N}_{103} - 15/2  \left( {\cal N}_{121} + {\cal N}_{301} \right) \nonumber \\
C_{42} &=& 45 ( {\cal N}_{202} - {\cal N}_{022} ) + 15/2 ( {\cal N}_{040} - {\cal N}_{400} ) \nonumber \\
C_{43} &=& 105 {\cal N}_{301} - 315 {\cal N}_{121} \nonumber \\
C_{44} &=& 105 ( {\cal N}_{400} + {\cal N}_{040} ) - 630 {\cal N}_{220} \nonumber
\end{eqnarray}
and for $S_{lm}$:
\begin{eqnarray}
S_{21} &=& 3 {\cal N}_{011} \nonumber \\
S_{22} &=& 6 {\cal N}_{110} \nonumber \\
S_{31} &=& 6 {\cal N}_{012} - 3/2 \left( {\cal N}_{030} + {\cal N}_{210} \right) \nonumber \\
S_{32} &=& 30 {\cal N}_{111} \nonumber \\
S_{33} &=& 45 {\cal N}_{210} - 15 {\cal N}_{030} \nonumber \\
S_{41} &=& 10 {\cal N}_{013} - 15/2 \left( {\cal N}_{031} + {\cal N}_{211} \right) \nonumber \\
S_{42} &=& 90 {\cal N}_{112} - 15 \left( {\cal N}_{130} + {\cal N}_{310} \right) \nonumber \\
S_{43} &=& 315 {\cal N}_{211} - 105 {\cal N}_{031} \nonumber \\
S_{44} &=& 420 ( {\cal N}_{310} - {\cal N}_{130} ) \nonumber
\end{eqnarray}

In order to solve the inverse problem, that consists of expressing 
the coefficients ${\cal N}_{ijk}$ as a function of $C_{lm}$ and $S_{lm}$,
we need to solve for ${\cal N}_{ijk}$ the linear system:
\begin{eqnarray}
C_{lm} &=& \gamma_{lm}^{ijk} {\cal N}_{ijk} \nonumber \\
S_{lm} &=& \sigma_{lm}^{ijk} {\cal N}_{ijk} \nonumber
\end{eqnarray}
where each of the two matrices $\gamma_{lm}^{ijk}$ and $\sigma_{lm}^{ijk}$ is real and two-dimensional (rectangular in general):
\begin{eqnarray}
\left[
\begin{array}{c}
C_{00} \\
C_{10} \\
\vdots \\
C_{lm} 
\end{array}
\right]
=
\left[
\begin{array}{cccc}
\gamma_{00}^{000} & \gamma_{00}^{100} & \cdots & \gamma_{00}^{ijk} \\
\gamma_{10}^{000} & \gamma_{10}^{100} & \cdots & \gamma_{10}^{ijk} \\
\vdots            &                   &        & \vdots            \\
\gamma_{lm}^{000} & \gamma_{lm}^{100} & \cdots & \gamma_{lm}^{ijk}
\end{array}
\right]
\left[
\begin{array}{c}
{\cal N}_{000} \\
{\cal N}_{100} \\
\vdots  \\
{\cal N}_{ijk}
\end{array}
\right]
\nonumber
\end{eqnarray}
We are allowed to split the solution of the inverse problem into the inversion
of two separate matrices because each one of the terms ${\cal N}_{ijk}$
is involved in either Eq.~(\ref{C_series_final}) or Eq.~(\ref{S_series_final}).
A close inspection of these equations in fact reveals that:
\begin{itemize}
\item both $C_{lm}$ and $S_{lm}$ contain only terms ${\cal N}_{ijk}$ where $i+j+k \equiv l$;
\item $C_{lm}$ contains only terms ${\cal N}_{ijk}$ where $j=2q+2\nu_y$ is an even integer;
\item $S_{lm}$ contains only terms ${\cal N}_{ijk}$ where $j=2q+1+2\nu_y$ is an odd integer.
\end{itemize}
If we consider only matrices $\gamma_{lm}^{ijk}$ and $\sigma_{lm}^{ijk}$ 
containing all and only terms of order $l=i+j+k$, 
we have that $\gamma_{lm}^{ijk}$ has dimensions $(l+1)\times{\lfloor (l+2)^2/4 \rfloor}$,
while $\sigma_{lm}^{ijk}$ has dimensions $l\times{\lfloor (l+1)^2/4 \rfloor}$. 
A total of $2l+1$ equations involving $2l+1$ spherical harmonics coefficients 
and $(l+1)(l+2)/2$ product of inertia coefficients.
As it is clear, the problem of expressing the ${\cal N}_{ijk}$ coefficients
in terms of the spherical harmonics coefficients is underdetermined, 
and has infinitely many solutions (if there are any).

\section{${\cal N}_{ijk}$ Coefficients for Homogeneous Bodies with Regular Shape\label{sec_precomputed}}

With the aim to provide more ready-to-use formulas,
we now compute the value of ${\cal N}_{ijk}$ for
a body having uniform mass distribution and
the shape of box, a cylinder, and a triaxial ellipsoid.
Because of these particular choices, the expression for ${\cal N}_{ijk}$
provided by Eq.~(\ref{eq_def_N}) simplifies as follows:
\begin{equation}
\displaystyle
{\cal N}_{ijk} =
\frac{\displaystyle \int_V \frac{x^i y^j z^k}{r_0^{i+j+k}} \delta(x,y,z) \ {\mathrm d}V}{\displaystyle \int_V \delta(x,y,z) \ {\mathrm d}V} =
\frac{1}{V r_0^{i+j+k}} \int_V x^i y^j z^k \ {\mathrm d}V
\label{eq_Nijk_omogeneous}
\end{equation}
Furthermore, because we choose the principal axes of the body as reference system, 
and the shape of each body is axis-symmetrical in each of the three principal directions,
the integral in Eq.~(\ref{eq_Nijk_omogeneous})
vanishes when any one of $i$, $j$, or $k$ is an odd integer.
So we will only consider expressions for ${\cal N}_{ijk}$
where $i$, $j$, and $k$ are even integers.
Expression analogous to those presented here are available in \cite{potential.book}.

\subsection{Box}

For a cartesian box with dimensions
$2a \times 2b \times 2c$ respectively along $x$, $y$, and $z$,
the volume is $V = 8 a b c$ and we have:
\begin{equation}
{\cal N}_{ijk} = 
\displaystyle
\frac{\displaystyle a^{i} b^{j} c^{k}}{\displaystyle r_0^{i+j+k}}
\frac{\displaystyle 1}{\displaystyle (i+1)(j+1)(k+1)} 
\nonumber
\end{equation}

\subsection{Cylinder}

For a cylinder with a circular base of radius $r$ parallel to the $x$-$y$ plane,
and height $2h$ along the $z$ axis, the volume is $V=2h \pi r^2$ and we have:
\begin{equation}
{\cal N}_{ijk} = 
\displaystyle
\frac{\displaystyle r^{i+j} h^{k}}{\displaystyle r_0^{i+j+k}}
\frac{1}{\displaystyle 2^{(i+j)/2} \left({(i+j)}/{2}+1 \right)! (k+1)} 
\prod_{p=1}^{i/2} (2p-1) 
\prod_{q=1}^{j/2} (2q-1)
\nonumber
\end{equation}

\subsection{Triaxial Ellipsoid}

For a triaxial ellipsoid with equation: $(x/a)^2+(y/b)^2+(z/c)^2 \leq 1$, 
the volume is $V=(4/3)\pi a b c$ and we have:
\begin{equation}
{\cal N}_{ijk} = 
\displaystyle
3
\frac{\displaystyle a^i b^j c^k}{\displaystyle r_0^{i+j+k}} 
\frac{\displaystyle \prod_{p=1}^{i/2} (2p-1) \prod_{q=1}^{j/2} (2q-1) \prod_{s=1}^{k/2} (2s-1)}{\displaystyle \prod_{u=1}^{(i+j+k)/2+2} (2u-1)}
\label{eq_triaxial_Nijk}
\end{equation}
The coefficients generated by Eq.~(\ref{eq_triaxial_Nijk}) up to the eighth order are:
\begin{eqnarray*}
{\cal N}_{000} &=& 1 \\
{\cal N}_{002} &=& 1/5    \left( c^2/r_0^2 \right) \\
{\cal N}_{004} &=& 3/35   \left( c^4/r_0^4 \right) \\
{\cal N}_{022} &=& 1/35   \left( b^2 c^2/r_0^4 \right) \\
{\cal N}_{006} &=& 1/21   \left( c^6/r_0^6 \right) \\
{\cal N}_{024} &=& 1/105  \left( b^2 c^4/r_0^6 \right) \\
{\cal N}_{222} &=& 1/315  \left( a^2 b^2 c^2/r_0^6 \right) \\
{\cal N}_{008} &=& 1/33   \left( c^8/r_0^8 \right) \\
{\cal N}_{026} &=& 1/231  \left( b^2 c^6/r_0^8 \right) \\
{\cal N}_{044} &=& 1/385  \left( b^4 c^4/r_0^8 \right) \\
{\cal N}_{224} &=& 1/1155 \left( a^2 b^2 c^4/r_0^8 \right) 
\end{eqnarray*}
Those not listed here can be easily obtained using the symmetry 
of Eq.~(\ref{eq_triaxial_Nijk}) for permutations of $ijk$.


\end{document}